%% file: main.tex
\documentclass{article}
\usepackage[T1]{fontenc} 
\usepackage[utf8]{inputenc} 
\usepackage{ismir,amsmath,url}
\usepackage{graphicx}
\usepackage{color, colortbl}
\usepackage{tabularx}       
\usepackage{multirow}
\usepackage{amsfonts}       
\usepackage{nicefrac}       
\usepackage{epigraph}
\usepackage{soul}
\usepackage[dvipsnames, table]{xcolor}
\usepackage{paralist}
\usepackage{caption}
\usepackage{subcaption}
\usepackage{pifont} 
\usepackage{wrapfig}
\usepackage{makecell}
\usepackage{array}
\usepackage[export]{adjustbox}
\usepackage{tablefootnote}

\usepackage{framed}
\usepackage{mathtools}
\usepackage{verbatim} 
\usepackage{enumerate}
\usepackage{bbm}
\usepackage{amsbsy}
\usepackage{float}
\usepackage[inline]{enumitem}

\usepackage[backend=biber,style=numeric-comp, maxnames=10, doi=false,isbn=false,url=false,eprint=false,date=year]{biblatex}
\AtEveryBibitem{%
  \clearfield{note}%
  \clearfield{editor}
    \clearname{editor}
}
\usepackage{lineno}

\usepackage{pifont}
\newcommand{\xmark}{\ding{55}}%

\usepackage{booktabs,arydshln}
\makeatletter
\def\adl@drawiv#1#2#3{%
        \hskip.5\tabcolsep
        \xleaders#3{#2.5\@tempdimb #1{1}#2.5\@tempdimb}%
                #2\z@ plus1fil minus1fil\relax
        \hskip.5\tabcolsep}
\newcommand{\cdashlinelr}[1]{%
  \noalign{\vskip\aboverulesep
           \global\let\@dashdrawstore\adl@draw
           \global\let\adl@draw\adl@drawiv}
  \cdashline{#1}
  \noalign{\global\let\adl@draw\@dashdrawstore
           \vskip\belowrulesep}}
\makeatother

\makeatletter
\renewcommand{\paragraph}{%
  \@startsection{paragraph}{4}%
  {\z@}{1ex \@plus 1ex \@minus .2ex}{-1em}%
  {\normalfont\normalsize\bfseries}%
}
\makeatother

\def\ourdata{MusicTextHQ}

\title{Augment, Drop \& Swap: \\ Improving Diversity in LLM Captions for Efficient Music-Text Representation Learning}

\threeauthors
  {Ilaria Manco} {Queen Mary University of London \\ {\tt \small i.manco@qmul.ac.uk}}
  {Justin Salamon} {Adobe Research \\ {\tt \small salamon@adobe.com}}
  {Oriol Nieto} {Adobe Research \\ {\tt \small onieto@adobe.com}}

\def\authorname{I. Manco, J. Salamon, and O. Nieto}

\usepackage[bookmarks=false,pdfauthor={\authorname},pdfsubject={\papersubject},hidelinks]{hyperref}

\begin{document}

\maketitle

\begin{abstract}
Audio-text contrastive models have become a powerful approach in music representation learning. Despite their empirical success, however, little is known about the influence of key design choices on the quality of music-text representations learnt through this framework.
In this work, we expose these design choices within the constraints of limited data and computation budgets, and establish a more solid understanding of their impact grounded in empirical observations along three axes: the choice of base encoders, the level of curation in training data, and the use of text augmentation. 
We find that data curation is the single most important factor for music-text contrastive training in resource-constrained scenarios. Motivated by this insight, we introduce two novel techniques, Augmented View Dropout and TextSwap, which increase the diversity and descriptiveness of text inputs seen in training. Through our experiments we demonstrate that these are effective at boosting performance across different pre-training regimes, model architectures, and downstream data distributions, without incurring higher computational costs or requiring additional training data.
\end{abstract}

\begin{figure*}[t]
\centering
\includegraphics[scale=0.3]{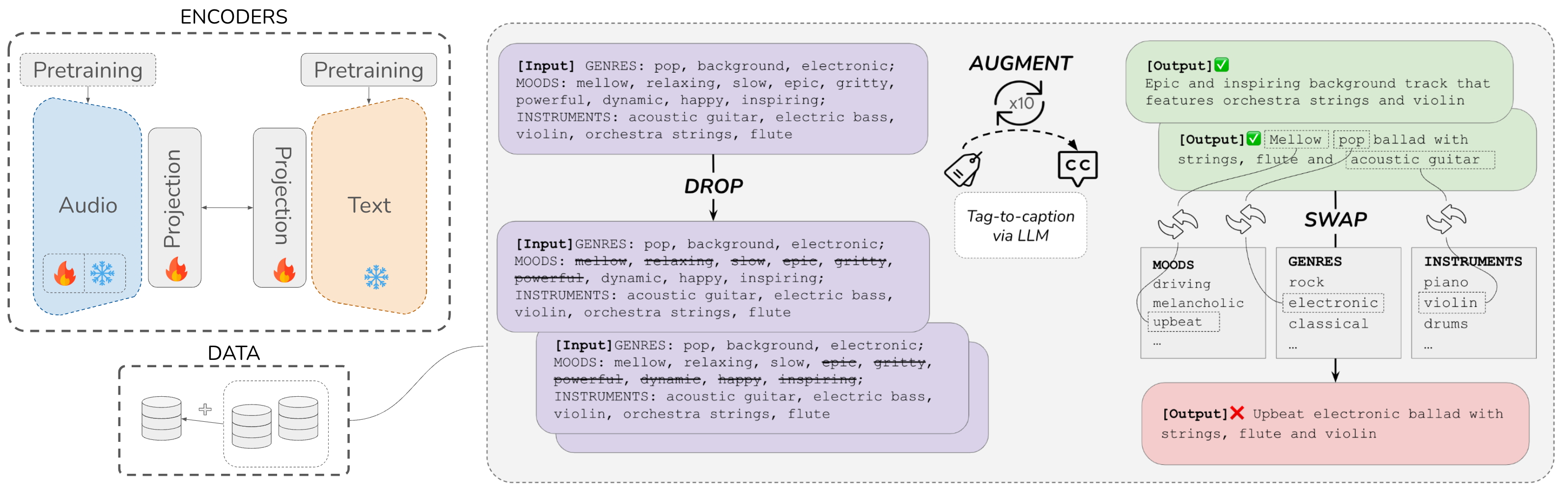}
\caption{\textbf{Overview} of our approach. We study the role of encoders and data in music-text learning and propose a text augmentation pipeline, \textit{Augment, Drop \& Swap}, to increase data diversity and introduce hard negatives during training.}
\label{fig:overview}
\end{figure*}

\section{Introduction}\label{sec:introduction}
Music-text embedding models have become a cornerstone of music information retrieval (MIR), facilitating core tasks that underpin music organisation and search, such as music tagging and cross-modal retrieval \cite{Manco2022, huang_mulan_2022, doh_toward_2022, Won2021, Won2021a}. At a high level, these are multimodal models that produce aligned audio-text representations by learning to project high-dimensional data from the audio and text modalities onto a lower-dimensional joint representation space whose structure encodes semantic similarity. The canonical learning framework to obtain such embeddings is dual-encoder multimodal contrastive learning, first popularised by CLIP \cite{Radford2021} in the image domain, and soon after adopted in most areas of machine perception, including audio \cite{elizalde_clap_2023, wu_large-scale_2023} and music processing \cite{Manco2022, huang_mulan_2022, doh_toward_2022}. 

Driven by the empirical success of this framework, a recent line of research has attempted to analyse its inner workings from a theoretical perspective \cite{nakada_understanding_2023, zhang_generalization_2023} or elucidate which aspects are most responsible for its effectiveness in visual models \cite{Zhai2021, you_learning_2022}.
However, within the audio domain, our understanding of multimodal contrastive learning remains limited \cite{doh_toward_2022}, with sparse effort into ablating design choices, or training data- and compute-efficient models. Among prior work that takes a step in this direction, the focus is mostly on comparing backbone models \cite{wu_large-scale_2023, huang_mulan_2022, doh_toward_2022}, but without considering other important factors such as model initialisation or training data. 
Additionally, audio-text learning poses specific challenges in the context of music, as the amount of data with aligned audio and text is typically orders of magnitude smaller than in other domains, where large-scale web-crawled data is commonplace. This makes transferring insights from other areas of representation learning particularly challenging.

In this paper we present a deep dive into music-text contrastive learning and its use in text-based music retrieval, adopting a practical perspective and thoroughly investigating the impact of major design choices. In particular, we study the problem of how to train this family of models under different resource-constrained scenarios (with respect to data and compute), and how to meaningfully evaluate them for real-world use. In brief, our contributions are as follows: \begin{enumerate*}[label=(\roman*)]
\item we systematically compare backbone encoders in parameter-efficient settings, and demonstrate that we can leverage this to enable multilingual support for the first time and without additional training data (Section \ref{sec:encoders}); 
\item we study the trade-off between training dataset size and quality, showing that the impact of data curation outweighs that of scale (Section \ref{sec:data}); 
\item building upon these findings, we propose a training recipe, \textit{Augment, Drop \& Swap} to construct more effective contrastive views (via Augmented View Dropout) and improve model robustness (via TextSwap) with no extra computational overhead (Section \ref{sec:ads}). Incorporating the proposed pipeline within variants of the music-text contrastive framework under different computational constraints, we show that this consistently improves over prior work, establishing a new state-of-the-art on three benchmark datasets. Finally, we conduct the first listening study to evaluate text-based music retrieval, further corroborating our automatic evaluations and underscoring the importance of accounting for distribution gaps when measuring performance.
\end{enumerate*}

\section{Studying the Design Space of Music-Text Embedding Models} \label{sec:background}
We explore two major factors in the design of music-text embedding models: architecture and data. While we acknowledge that there are others, such as training procedure, and alternative designs, we choose to restrict our focus exclusively to these two axes and to dual-encoder models, due to their predominance in the field. In the rest of the paper, we always refer to this family of models when discussing \textit{music-text embeddings} or \textit{music-text models}, and interchangeably use the terms \textit{text} and \textit{language}.

The typical design of a music-text embedding model consists of the following components: two modality-specific base encoders which separately process inputs of the text and audio modality to an intermediate representation space; a fusion or projection module responsible for mapping the intermediate representations to the shared embedding space; and a contrastive loss, through which the model parameters are optimised to encode semantically related audio and text inputs within the same neighbourhood of the embedding space, while pushing apart unrelated items. We provide an overview of this design in Figure \ref{fig:overview}. While prior works have converged towards standard choices for the last two components, it remains unclear how to reliably choose unimodal encoders among several existing options. We look at this in Section \ref{sec:encoders}, before discussing the role of training data in Section \ref{sec:data}.

\subsection{Our experimental approach}\label{sec:experimental_setup} Before delineating our areas of focus, we outline here the standard experimental setup used in our experiments.

\paragraph*{Projection module} We design our experiments to compare variations of the dual-encoder contrastive architecture described above, varying several components, but keeping two fixed throughout: the projection module and the loss. Similarly to \cite{mckee_language-guided_2023, suris_its_2022}, we adopt a two-head, two-layer Transformer as our projection module. From a sequence of 256-dimensional embeddings produced by each projection head, we employ the \texttt{[CLS]} token embedding as the global representation for each branch. For ease of reference, we denote this model architecture by DuET-MC (\textbf{Du}al-\textbf{E}ncoder \textbf{T}ext-\textbf{M}usic \textbf{C}ontrastive).

\paragraph*{Training}
We optimise our network via the multimodal formulation of the InfoNCE loss \cite{VanDenOord2018}, using cosine similarity between the \textit{l2}-normalised projection embeddings from the audio and text branch as our scoring function, and a temperature parameter of 0.03. As part of our training procedure, we use the Adam optimizer with decoupled weight decay of 0.05, varying our learning rate through a cosine decay schedule from its peak value of 1e-3, after a linear warm-up of 5 epochs. We train on 8 A100 NVIDIA GPUs, with an effective batch size of 1024 or 2048 based on memory requirements, for a maximum of 100 epochs, with early stopping based on the validation loss.
Unless otherwise specified, our default training data is a corpus of licensed instrumental music with high-quality, manually curated genre, mood, and instrument tags, which we refer to as \ourdata. For training, we select a subset totalling a duration of 100 hours, and augment tags into captions following our data augmentation strategy described in Section \ref{sec:data}.

\subsection{Evaluation}
We evaluate all our models on text-based music retrieval, as this represents the most prominent task for music-text embedding models and has been shown to correlate to performance on other tasks \cite{Manco2022, huang_mulan_2022}. 
Retrieval is performed by ranking all audio clips in the dataset by decreasing cosine similarity of their embedding with the embedding of a text query. From this, we compute Recall@$k$ (R@$k$), the average number of times the target appears within the top-$k$ retrieved items, and Median Rank (MR). To normalise performance scores by the different dataset sizes, we repeat this procedure on random subsets of 500 items, and report the average value for each metric. When reporting a single metric, we always refer to R@10.

\paragraph*{Datasets} In order to robustly measure performance across our experiments, we adopt a multi-dataset evaluation suite comprising three public datasets containing audio tracks paired with human-written captions: YT8M-MusicTextClips (MTC) \cite{mckee_language-guided_2023}, MusicCaps (MC) \cite{agostinelli_musiclm_2023} and Song Describer (SDD) \cite{manco_song_2023}. These all represent out-of-distribution data (Table \ref{tab:datasets}), with different degrees and types of distribution shifts in both the audio and text modality. For example, MTC and MC both contain 10-second audio clips from YouTube videos, but they differ significantly in their captions, with respect to content, descriptiveness and even text length \cite{manco_song_2023}. Audio in the SDD consists instead of music recordings from the music platform Jamendo \cite{Bogdanov2019}, while captions describe much longer audio segments.

\begin{table}[t]
\small
\centering
\begin{tabular}{p{3cm}ccccc}\toprule
Dataset & Hours* & Tags & Captions \\
\midrule
\textit{\small{Training}} \\
LP-MusicCaps \cite{doh_lp-musiccaps_2023} (\texttt{A}) & 50 & Human & Synthetic\\
\ourdata{} (\texttt{B}) & 100  & Human & Synthetic \\
YT8M-MV \cite{abu-el-haija_youtube-8m_2016} (\texttt{C}) & 270 & Synthetic & Synthetic \\
\cdashlinelr{1-5}
\textit{\small{Evaluation}} \\
YT8M-MTC \cite{mckee_language-guided_2023} & 8 & - & Human \\
MusicCaps \cite{agostinelli_musiclm_2023} & 8 & - & Human\\
Song Describer \cite{manco_song_2023} & 2 & - & Human\\
\bottomrule
\end{tabular}
\caption{\textbf{Overview of the datasets} used in our experiments. *Hours denotes the audio duration used in training.}
\vspace{1em}
\label{tab:datasets}
\end{table}

\begin{table}[t]
\small
\centering
\begin{tabular}{llc}\toprule
Encoder & \# Params  & Model version \\
\midrule
\textit{\small{Audio}} \\
HTS-AT \cite{chen_hts-at_2022} & 30M & \texttt{AudioSet}\tablefootnote{We use the \texttt{HTSAT\_AudioSet\_Saved\_6} checkpoint of HTS-AT trained on AudioSet from the official repository.}  \\
MERT \cite{li_mert_2023}  & 330M & \texttt{MERT-v1-330M} \\
\cdashlinelr{1-3}
\textit{\small{Text}} \\
RoBERTa \cite{liu_roberta_2019} & 125M & \texttt{roberta-base} \\
CLIP-T \cite{Radford2021} & 151M & \texttt{clip-vit-base-patch32} \\
T5 \cite{Raffel2019} & 11.3B & \texttt{flan-t5-xx} \\
mT5 \cite{xue_mt5_2021} & 13B & \texttt{mt5-xx}  \\ 
\bottomrule
\end{tabular}
\caption{\textbf{Audio and text encoders} we compare in our experiments on the impact of encoder backbones (Section \ref{sec:encoders}).}
\label{tab:encoders}
\end{table}

\section{The role of encoder backbones}\label{sec:encoders}
We experiment with two audio encoders, HTS-AT \cite{chen_hts-at_2022} and MERT \cite{li_mert_2023}, and three text encoders, RoBERTa \cite{liu_roberta_2019}, the text encoder from CLIP \cite{Radford2021} (CLIP-T), T5 \cite{Raffel2019} and mT5 \cite{xue_mt5_2021}. We choose these either because they represent the state of the art in their respective tasks, or because they have been previously used in contrastive audio-text learning, thus allowing for direct comparison with prior work.

\subsection{Encoders: initialization and freezing}\label{sec:pretraining}
In this set of experiments, our goal is to study parameter-efficient configurations of existing audio and text encoders, training only a subset of the model weights. The motivation for exploring this setting is threefold: freezing part of the model lowers the memory budget and training time, it avoids catastrophic forgetting \cite{mehta_empirical_2023}, and it reduces the risk of overfitting in data-constrained scenarios. To fulfil these requirements, we do not consider end-to-end finetuning, and instead focus on leveraging pre-training, locking the audio and text encoders based on their parameter size. Specifically, we keep all text encoders frozen, as these all count over 100M parameters, as shown in Table \ref{tab:encoders}, and only train the full audio encoder, both with and without general-audio pre-training when using HTS-AS, due to its smaller size.
When using MERT, we keep the encoder frozen but train a learnable aggregator over the hidden states of each layer, implemented as a 1D convolutional layer, to obtain audio representations that capture the different levels of abstraction encoded at different depths of the network \cite{li_mert_2023}.

\begin{figure}[t]
\centering
 \includegraphics[scale=0.5]{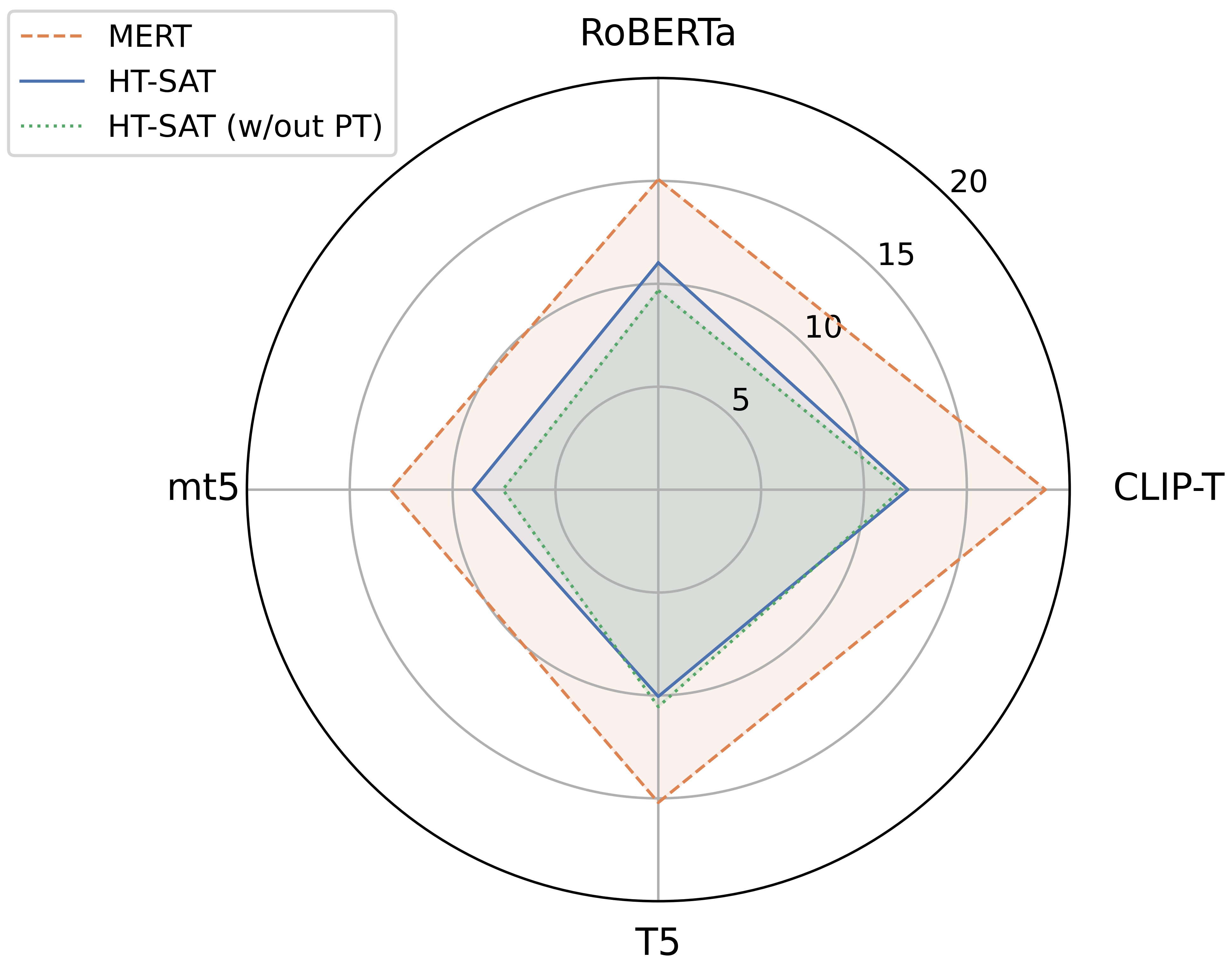}
 \caption{\textbf{Retrieval performance (R@10) of different combinations of audio and text encoders} compared through the lens of our DuET-MC framework.}
 \label{fig:encoders}
\end{figure}

\paragraph*{Results}
From Figure \ref{fig:encoders} we first observe that, under the constraints described above, the overall best configuration is given by MERT and CLIP-T. We attribute this to two main reasons: with regards to the audio branch, the superior performance exhibited by MERT suggests that a larger model capacity and stronger music prior may be beneficial to music-text alignment; with regards to the text branch, while all encoders are characterised by large-scale pre-training, CLIP-T stands out as the only model with multimodal capabilities. Although this is somewhat surprising, as CLIP is pre-trained on image-text pairs, we note that prior work has also shown that it can be successfully transferred to the audio and music domains \cite{wu_wav2clip_2022, mckee_language-guided_2023, chen_learning_2022, yang_diffsound_2023}. Secondly, when using MERT with any of the text encoders considered, we find that we can train less than 1\% of the total amount of weights ($\sim3$M making up the projection and aggregation layers) without loss of performance compared to current state-of-the-art models (shown later in Table \ref{tab:ablations}). This demonstrates that we can successfully align locked text representations to the audio modality through light-weight music-text contrastive learning, confirming that our encoder locking strategy is effective when leveraging powerful music-specific pre-training, in line with similar findings in the visual domain \cite{Zhai2021}. With regards to the audio branch initialization, comparing the two variants of HTS-AT, we find that general-purpose audio pre-training can give a slight advantage over training from scratch, but this benefit is not consistent across the different text encoders HTS-AT is paired with. In the rest of the paper we fix the encoder configuration to locked MERT + locked CLIP-T in all experiments, unless otherwise specified.

\subsection{Supporting retrieval in multiple languages}\label{sec:multilingual}
Due to a lack of data in different languages, music-text modelling has so far exclusively focussed on English. Real-world applications for music-text embeddings, however, can greatly benefit from the support of multiple languages. To address this limitation, we explore the use of pre-trained locked encoders, similarly to Section \ref{sec:pretraining}, this time adopting mT5 \cite{xue_mt5_2021}, a multilingual text-to-text Transformer model, as our text encoder. To evaluate multilingual performance, we choose a subset of four languages, German, French, Italian and Spanish, and translate our evaluation datasets via \texttt{GPT3.5-turbo} \cite{Brown2020}. In Table \ref{tab:multilingual}, we show that this approach provides a viable solution to text-based retrieval in multiple languages while using only English text paired with music in training and with only a minor drop in performance compared to English. 

\begin{table}[t]
\small
\centering
\begin{tabular}{cccc}\toprule
 \multirow{2}{*}{Language} & \multicolumn{3}{c}{R@10} \\
 \cmidrule(rl){2-4} 
&  YT8M-MTC   & MusicCaps   & Song Describer   \\ 
\midrule
English & 10.43  & \textbf{16.00} & \textbf{19.00} \\
German  & 9.90  & 13.28  & 18.00 \\
French  & 11.71  & 12.32  & 15.40 \\
Italian  & 10.43  & 13.68  & 15.80 \\
Spanish  & \textbf{11.60}  & 13.48 & 18.40 \\
\bottomrule
\end{tabular}
\caption{\textbf{Multilingual retrieval performance}.}
\label{tab:multilingual}
\end{table}

\section{The role of training data}\label{sec:data}
Having established best practices with respect to choosing audio and text backbones, we now shift our attention to the training data. As widely acknowledged in the literature \cite{doh_toward_2022, doh_lp-musiccaps_2023, Manco2022}, a major limitation in training music-language models is the lack of large public datasets with paired audio-text data. To circumvent this issue, a number of works have proposed to employ large language models to augment text data more commonly found in music datasets, such as categorical labels, metadata and tags, into full natural language sentences, corresponding to \textit{pseudo-captions} \cite{mckee_language-guided_2023, doh_lp-musiccaps_2023, gardner_llark_2023}. In the next section we present our investigation of the impact of tag-to-caption augmentation.

\subsection{Tag-to-caption augmentation via LLMs}\label{sec:augmentation}
Following \cite{mckee_language-guided_2023}, we leverage the in-context learning ability of LLMs via few-shot prompting, and adopt a similar approach to augment tags into captions for our training dataset \ourdata{}. For this, we use BLOOM-176B \cite{scao_bloom_2023}, a competitive open-access LLM trained on responsibly sourced data. Differently from \cite{mckee_language-guided_2023}, we do not employ synthetic tags, but use tags provided by expert annotators.
We compare this to training on LP-MusicCaps-MTT \cite{doh_lp-musiccaps_2023} (LP-MusicCaps for short), a dataset obtained via a similar approach, where tags from the MagnaTagATune \cite{Law2009} dataset are augmented into captions via \texttt{GPT3.5-turbo}. To measure the impact of tag-to-caption augmentation, we train three variants of our model on each dataset, varying $p_{cap}$, the probability of selecting captions over tags as the text input for each training pair. 

\begin{figure}[t]
\centering
 \includegraphics[scale=0.38]{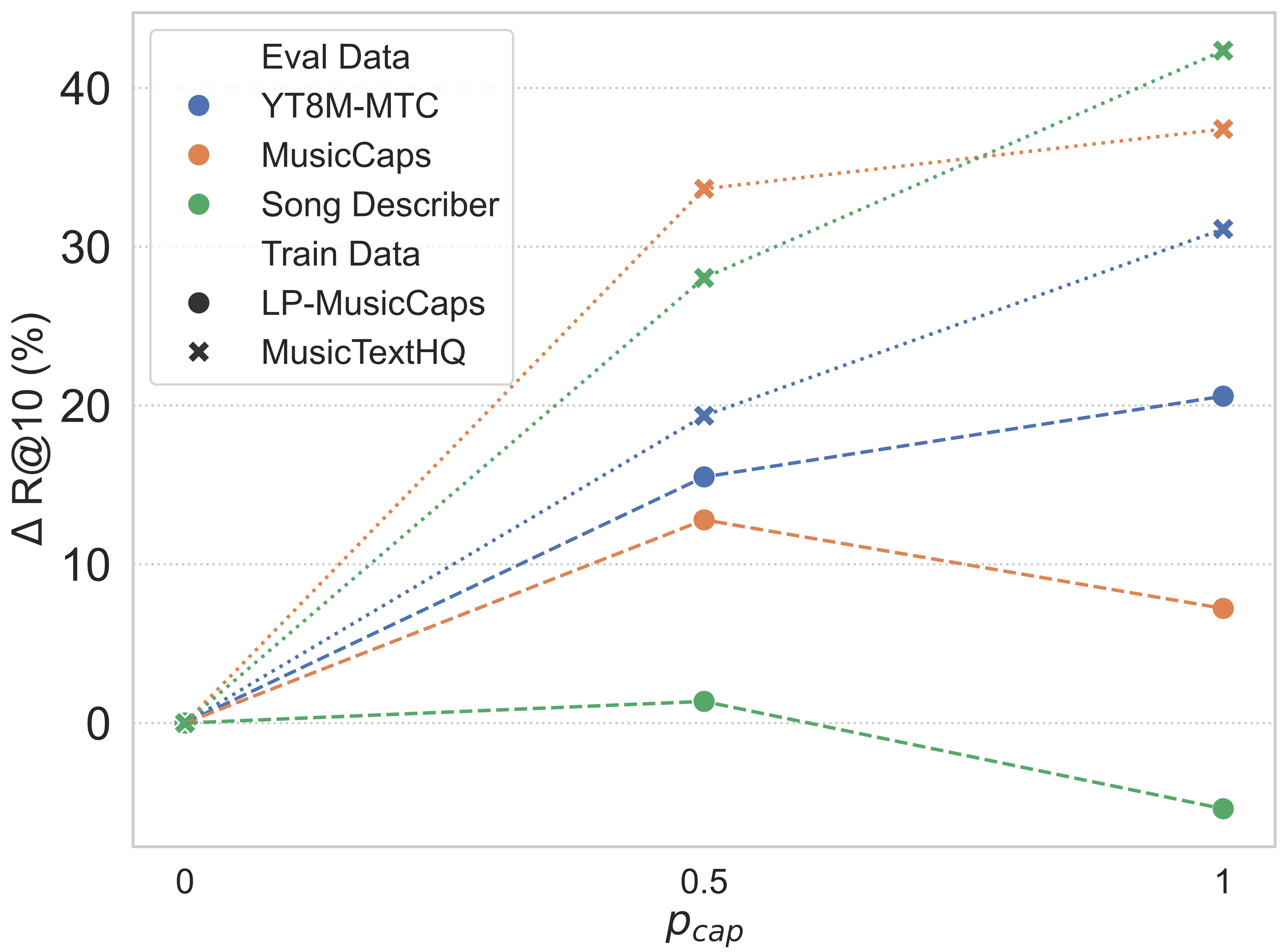}
 \caption{\textbf{The effect of varying $\boldsymbol{p_{cap}}$}, the probability of swapping tags with captions. On the y-axis, we show the relative change in performance compared to $p_{cap}=0$.}
 \label{fig:tag_vs_cap}
\vspace{-0.6em}
\end{figure}

\paragraph*{Results} Results are shown in Figure \ref{fig:tag_vs_cap}, where we compare the effect of gradually shifting from tags to captions in the two training datasets considered. We first note that introducing tag-to-caption augmentation for at least a portion of the training data ($p_{cap}=0.5$) leads to an improvement regardless of training dataset. Interestingly, unlike in \ourdata{}, this trend does not extend to the scenario where we replace all text inputs with pseudo-captions ($p_{cap}=1$) in LP-MusicCaps. In this case, we observe instead a slight degradation in performance on two of the evaluation datasets, compared to using only tags, or using captions half of the time. We posit that this divergence may be due to a gap in label quality between the two training sets, as pseudo-captions in LP-MusicCaps are generated based on sparse labels, with 50\% of the items in the dataset paired to only three tags or less, and in many cases without being balanced across categories. Intuitively, this is likely to result in non-descript, or even inaccurate captions, as the LLM generation will be more prone to hallucinations and may therefore deviate substantially from the audio content. In contrast, \ourdata{} provides strong grounding to the audio content, with multiple expert-provided tags per category (often three or more for \textit{each} tag category). From this, we conclude that, while LLM-enabled text augmentation can provide a valuable strategy for enriching training data, it is not a substitute for adequate data curation, but rather a supplement. This is an important observation, as prior work has also found that specificity in captions is instrumental to effective multimodal contrastive learning \cite{xue_understanding_2023, santurkar_is_2022}. Since LLM-based augmentation, being bounded by the information content in the source data, cannot increase specificity, our results highlight an often overlooked shortcoming of synthetic text. 

\subsection{Training data: size vs quality}\label{sec:quality_vs_size}
Next, we ask whether simply increasing dataset size can emphasise the benefits of tag-to-caption augmentation. To scale up the size of our training data, we include YT8M-MV, a subset of the YouTube8M dataset \cite{abu-el-haija_youtube-8m_2016} tagged as \textit{music video}, as an additional dataset to our training pool. For this, we follow \cite{mckee_language-guided_2023} and employ tags from an automatic music tagger and pseudo-captions generated following the same procedure described in Section \ref{sec:augmentation}. For simplicity, we refer to LP-MusicCaps, MusicTextHQ and YT8M-MV as \texttt{Dataset\_A} (or simply \texttt{A}), \texttt{Dataset\_B} (\texttt{B}) and \texttt{Dataset\_C} (\texttt{C}), ordered by size as shown in Table \ref{tab:datasets}. We also consider combining the two biggest datasets (\texttt{B} + \texttt{C}) and all three together (\texttt{A} + \texttt{B} + \texttt{C}). We note that each dataset differs not only in size, but also in audio and label quality.

\begin{figure}[t]
\centering
 \includegraphics[scale=0.48]{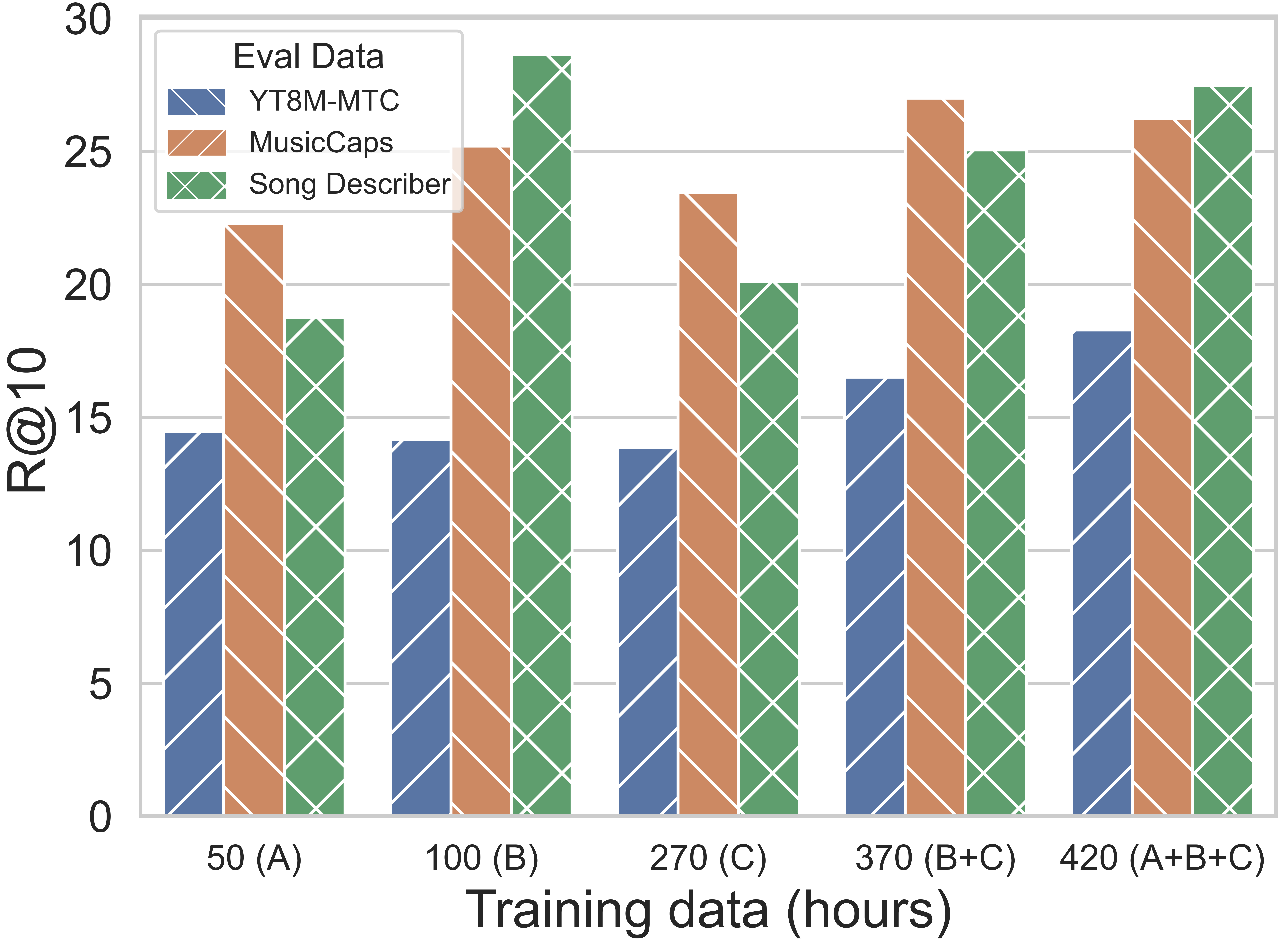}
 \caption{Retrieval performance across models trained on \textbf{datasets that differ in size and annotation quality}.}
 \label{fig:datasets}
   \vspace{-0.6em}
\end{figure}

\paragraph*{Results} In Figure \ref{fig:datasets} we showcase results from training on the datasets described above. Notably, we find that scaling dataset size does not consistently result in an improvement, signalling that the gap in quality between datasets can eclipse their size difference. Although we observe that combining all datasets yields better performance, likely due to overall increased diversity in the training data, the difference is not proportionate to the rise in training cost necessary to scale up. Instead, our results underscore the importance of data curation as a more efficient way of boosting performance, confirming that constructing a subset of highly curated examples, with descriptive and accurate captions, more positively contributes to learning in the contrastive setting \cite{santurkar_is_2022}.

\section{Improving diversity via text augmentations}\label{sec:ads}
Having established that augmenting high-quality tags into captions offers a useful and inexpensive strategy to enrich training data, we explore this further and propose two augmentation-based techniques aimed at increasing data diversity and model robustness.

\subsection{Augment, Drop \& Swap}
\paragraph*{Augmented View Dropout} 
First, building upon the tag-to-caption strategy described in Section \ref{sec:augmentation}, we explore text augmentation with the goal of constructing more effective views for contrastive learning, following the principle that optimal views should minimise mutual information between paired items while retaining a high degree of semantic alignment \cite{tian_what_2020}. To this end, we propose \textit{Augmented View Dropout}, where, for each item in our dataset, we randomly sample a subset of the tags, balanced by category (genre, mood, instrumentation) and produce a set of 10 different captions. Each can be thought of as a complementary, but partial view of the associated music track, as we mask a subset of all the ground-truth tags to produce each view. At training time, views are randomly sampled, effectively resulting in a further form of data augmentation. 

\paragraph*{Hard negatives via TextSwap}
Finally, we tackle another important challenge in contrastive learning, hard negative sampling, and propose to also address this through the lens of text augmentation, via a technique which we call \textit{TextSwap}. In order to increase the rate of hard negatives beyond the natural rate found in the dataset, we create partially perturbed versions of the captions by stochastically swapping genre, mood or instrument keywords with alternative descriptors from a predefined dictionary (e.g. ``a mellow \textit{pop} track'' becomes ``a mellow \textit{hip-hop} track''). During training, for each positive pair, we then select a random subset of the negative captions in a batch and replace them with hard negatives by applying TextSwap once per descriptor category. This is illustrated in Figure \ref{fig:overview}, where we provide a visual guide for the full \textit{Augment, Drop \& Swap} pipeline. We hypothesise that the presence of hard negatives is particularly critical in later stages of training, once the model has already acquired basic features, and learning on ``easy'' negatives has saturated. Based on this, we follow a curriculum learning approach and linearly increase the probability of applying TextSwap from 0 to 15\% over the course of 20 epochs, after a warm-up period of 5. 

\input{tables/ablations}

\subsection{Experiments}\label{sec:ads_experiments}
\paragraph*{Ablations} In this set of experiments we examine the effect of each of the three components in our augmentation pipeline: tag-to-caption augmentation, Augmented View Dropout and TextSwap. We look at two scenarios: one where we want to measure their contribution in training two variants of our parameter-efficient DuET-MC framework, each with different degrees of audio pre-training and finetuning and with locked text encoders, and one where we relax our computational requirements and explore whether our proposed method can be usefully applied in finetuning a general purpose audio-text embedding model (CLAP \cite{wu_large-scale_2023}), with limited paired music data. 

\paragraph*{Results} We present our ablations on the proposed pipeline in Table \ref{tab:ablations}, where we also compare to two audio-text contrastive baselines, CLAP \cite{wu_large-scale_2023} and TTMR \cite{doh_toward_2022}, trained on general-purpose audio and music respectively. 
The table displays three different settings to which we apply our proposed pipeline: (1) training the audio encoder from scratch (shown in the HTS-AT + CLIP-T configuration), (2) training only 1\% of the parameters in our locked audio-text encoder (MERT + CLIP-T), and (3) fine-tuning the full model on music, following general audio-text pre-training (CLAP-FT). From this, we observe that, while the vanilla version of DuET-MC (trained only on tags) exhibits at best comparable performance to the baselines, each additional component in our pipeline lifts performance across all model configurations, pre-training regimes and finetuning strategies. Among these, tag-to-caption augmentation and Augmented View Dropout emerge as the most influential, while the benefits of TextSwap are more prominent for model configurations where encoders have higher levels of pre-training, hinting at the necessity of increasing the complexity of negatives later in training. This suggests that our \textit{Augment, Drop \& Swap} recipe provides a data-efficient strategy to improve music-text modelling under a variety of model configurations, at no additional computational cost. Importantly, this trend generalises across evaluation datasets, suggesting that it is beneficial to model robustness, and demonstrates that the lack of large-scale paired data in the music domain can be alleviated through augmentation-based techniques which enhance data quality instead of quantity.
Finally, comparing retrieval scores of different family of models (TTMR, CLAP and DuET-MC), we note consistent differences between datasets, with CLAP-based models invariably showing a significant jump in performance on the MusicCaps dataset compared to MTC and SDD. We hypothesise that this may be a result of in-distribution bias, since there are several instances of non-music or noisy, low-quality recordings in MC. Since CLAP is trained to recognise everyday sounds, this points at a smaller shift from its training distribution, compared to SDD and MTC, which are exclusively composed of music recordings. We posit that further mismatches in the training and test distributions exist along the text dimension and investigate this through human evaluation.

\paragraph*{Are metrics aligned with human preference?} \label{sec:human_eval}
We recruit 35 participants to evaluate DuET-MC, CLAP and TTMR in a head-to-head pairwise comparison. Participants are presented with up to 24 text prompts, where each is a caption taken from one of the three evaluation datasets, and are asked to choose which one of two music tracks best aligns to the description. Through this qualitative evaluation, we find that DuET-MC does substantially better than TTMR, losing against it only 30.9\% of the times, and largely mirroring our findings from Section \ref{sec:ads_experiments}. Surprisingly, the win and tie rate vs CLAP drops instead to 37.3\% and 38.5\% respectively. Looking at the breakdown of scores by dataset, this advantage in CLAP is predominantly observed on MC and MTC, while DuET-MC outperforms CLAP on SDD. Interestingly, DuET-MC is preferred or considered equivalent to the ground truth 38.9\% of the times on MTC compared to 15.4 and 17.9\% on the other two datasets. This points to significant differences in the level of alignment between caption and audio in the different datasets, signalling that evaluating on several datasets is paramount to understanding real-world performance. Additionally, it leads to an observation that complements our automatic evaluation in Table \ref{tab:ablations}: the discrepancy between DuET-MC's performance on MTC compared to MC and SDD may be ascribed to a higher degree of \textit{vagueness} in MTC captions, which, as revealed through our qualitative evaluation, admit instead alternative matching tracks to those in the ground truth.

\section{Conclusions}
In this work we presented \textit{Augment, Drop \& Swap}, a training recipe for efficient music-text representation learning informed by our findings on training music-text contrastive models in resource-constrained scenarios. Through our experiments, we provide a practical guide to this family of models, and foreground their real-world use by focusing on multilingual support, computationally efficient techniques, and cross-dataset evaluation. Showing that data curation has a significant effect at modest data scales, we design each step in our pipeline to tackle specific aspects of the text used in training, such as descriptiveness and specificity, via data augmentations, leading to views that are more effective in multimodal contrastive learning. Through automatic and qualitative evaluations, we show the usefulness of our approach and reveal insights on the relation between measured performance and distribution shifts in the test data.

\printbibliography[title=References, heading=bibnumbered]

\end{document}

%% file: tables/ablations.tex
\begin{table*}[t]
\small
\centering
    \begin{tabular}{lccccccccccc}
    \toprule
 \multirow{2}{*}{Model} & \multirow{2}{1.3cm}{\centering Tag-to-caption} & \multirow{2}{1.8cm}{\centering Augmented View Dropout} & \multirow{2}{1.2cm}{\centering TextSwap} & \multicolumn{2}{c}{YT8M-MTC}   & \multicolumn{2}{c}{MusicCaps}   & \multicolumn{2}{c}{Song Describer}  & \multirow{2}{1.1cm}{\centering Avg R@10 $\uparrow$} \\
 \cmidrule(rl){5-6}  \cmidrule(rl){7-8}   \cmidrule(rl){9-10} 
&    &   &  & R@10 $\uparrow$ & MR $\downarrow$ & R@10 $\uparrow$ & MR $\downarrow$ & R@10 $\uparrow$ & MR $\downarrow$  \\ 
\midrule
\textit{Baselines} \\
CLAP \cite{wu_large-scale_2023} & - & - & - & \cellcolor{white} 11.9 & \cellcolor{white} 80 & \cellcolor{white} 40.3* & \cellcolor{white} 17* & \cellcolor{white} 19.8 & \cellcolor{white} 53 & \cellcolor{white} 24.0*\\
TTMR \cite{doh_toward_2022} & - & - & -  & \cellcolor{white} 11.6 & \cellcolor{white} 79 & \cellcolor{white} 9.6 & \cellcolor{white} 115 & \cellcolor{white} 16.5 & \cellcolor{white} 57 & \cellcolor{white} 12.6 \\
\cdashlinelr{1-11}

\multirow{4}{1.5cm}{DuET-MC (HTS-AT + CLIP-T)}  &  \xmark  & \xmark & \xmark & \cellcolor{white}  8.5 & \cellcolor{white} 103 & \cellcolor{white} 12.2 & \cellcolor{white} 82 & \cellcolor{white} 15.3 & \cellcolor{white} 53 & \cellcolor{white} 12.0\\ 

& \checkmark  & \xmark & \xmark &  \cellcolor{green!10} 8.0 & \cellcolor{white} 104 & \cellcolor{green!10} 13.4 &  \cellcolor{green!10} 76 & \cellcolor{white} 14.1 & \cellcolor{white} 57 & \cellcolor{white} 11.8 \\ 

& \checkmark & \checkmark & \xmark & \cellcolor{green!10} \underline{9.4}  & \cellcolor{green!10} \underline{93} & \cellcolor{green!10} 15.1 &  \cellcolor{green!10} \underline{65} & \cellcolor{green!10} \underline{19.6} &  \cellcolor{green!10} 49 & \cellcolor{green!10} \underline{14.7} \\ 

& \checkmark  & \checkmark &  \checkmark & \cellcolor{white} \underline{9.4} & \cellcolor{white} \underline{93} & \cellcolor{green!10} \underline{15.8} &  \cellcolor{white} 66 & \cellcolor{white} 17.4 &  \cellcolor{green!10} \underline{48} & \cellcolor{white} 14.2\\ 

\cdashlinelr{1-11}
\multirow{4}{1.5cm}{DuET-MC (MERT + CLIP-T)}  &  \xmark  & \xmark & \xmark & \cellcolor{white}  10.8 & \cellcolor{white} 82 & \cellcolor{white} 18.3 & \cellcolor{white} 56 & \cellcolor{white} 20.2 & \cellcolor{white} 45 & \cellcolor{white} 16.4 \\ 

& \checkmark  & \xmark & \xmark &  \cellcolor{green!10} 11.7 & \cellcolor{green!10} 69 & \cellcolor{green!10} 21.3 &  \cellcolor{green!10} 41 & \cellcolor{green!10} 23.4 &  \cellcolor{green!10} 36 & \cellcolor{green!10} 18.8 \\ 

& \checkmark & \checkmark & \xmark & \cellcolor{green!10} 13.4 & \cellcolor{green!10}  65 & \cellcolor{green!10} \underline{24.9} &  \cellcolor{green!10} 36 &  \cellcolor{green!10} \textbf{\underline{27.7}} & \cellcolor{green!10} 32 & \cellcolor{green!10} 22.0 \\ 

& \checkmark  & \checkmark &  \checkmark & \cellcolor{green!10} \underline{14.5} & \cellcolor{green!10} \underline{62} & 24.6 &  \cellcolor{green!10} \underline{34} &  \cellcolor{white} 27.3 &  \cellcolor{green!10} \textbf{\underline{29}} & \cellcolor{green!10} \underline{22.1} \\ 

\cdashlinelr{1-11}
\multirow{4}{*}{CLAP-FT}  &  \xmark  & \xmark & \xmark & \cellcolor{white}  14.2 & \cellcolor{white} 63 & \cellcolor{white} 38.8* & \cellcolor{white} 18* &  
\cellcolor{white} 20.8 & \cellcolor{white} 38 & \cellcolor{white} 24.6* \\ 
 
& \checkmark  & \xmark & \xmark & \cellcolor{green!10}  14.6 & \cellcolor{green!10} 61 & \cellcolor{green!10} 42.3* &  \cellcolor{green!10} 15* &  \cellcolor{green!10} 23.5 &  \cellcolor{green!10} 34 & \cellcolor{green!10} 26.8* \\ 

& \checkmark & \checkmark & \xmark & \cellcolor{green!10} \textbf{\underline{16.3}} & \cellcolor{green!10} \textbf{\underline{55}} & \cellcolor{white}  41.6* & \cellcolor{white} 16* &   \cellcolor{green!10} 24.5 & 36 & \cellcolor{green!10} 27.3* \\  

& \checkmark  & \checkmark &  \checkmark & \cellcolor{white} 15.7 & \cellcolor{white} 57 &  \cellcolor{green!10} \textbf{\underline{43.5}}* &  \cellcolor{green!10} \textbf{\underline{14}}* &  \cellcolor{green!10} \underline{26.3}  &  \cellcolor{green!10} \underline{31} & \cellcolor{green!10} \textbf{\underline{28.5}}* \\ 

\bottomrule
\end{tabular}
\caption{\textbf{Ablations}. For each model, subsequent rows show the effect of introducing an additional step in our proposed \textit{Augment, Drop \& Swap} pipeline. We highlight best results for each model (underlined) and amongst all models (bold). * denotes values that may be inflated due to in-distribution bias.}
\label{tab:ablations}
\end{table*}